\documentclass[aps,prl,showpacs,floatfix,superscriptaddress,twocolumn,longbibliography,nofootinbib]{revtex4-1}

\usepackage{amsmath,amssymb,physics}
\usepackage[toc,page]{appendix}
\usepackage{verbatim,graphicx,mathtools}
\usepackage{txfonts}
\usepackage{color}
\usepackage[all]{xy}
\usepackage{color}
\usepackage[all]{xy}
\usepackage{subfigure}

\newcommand{\be}{\begin{eqnarray}}
\newcommand{\ee}{\end{eqnarray}}

\newcommand{\bmat}{\left ( \begin{array}{cc} }
	\newcommand{\emat}{\end{array} \right ) }

\newcommand{\beq}{\begin{equation}}
\newcommand{\beqs}{\begin{equation*}}
\newcommand{\eeq}{\end{equation}}
\newcommand{\eeqs}{\end{equation*}}

\allowdisplaybreaks

\begin{document}

\author{Antonio M. Garc\'\i a-Garc\'\i a}
\affiliation{Shanghai Center for Complex Physics, 
	Department of Physics and Astronomy, Shanghai Jiao Tong
	University, Shanghai 200240, China}
\email{amgg@sjtu.edu.cn}
\author{Bruno Loureiro}
\affiliation{IdePHICS lab, Ecole F\'ed\'erale Polytechnique de Lausanne, Switzerland}
\email{brloureiro@gmail.com}
\author{Aurelio Romero-Berm\'udez}
\affiliation{Institute Lorentz for Theoretical Physics, Leiden University, Leiden, The Netherlands}
\email{aurelio.romero.bermudez@gmail.com}
\author{Masaki Tezuka}
\affiliation{Department of Physics, Kyoto University, Kyoto 606-8502, Japan}
\email{tezuka@scphys.kyoto-u.ac.jp}

\textbf{Reply to Comment on ``Chaotic-Integrable Transition in the Sachdev-Ye-Kitaev Model''}
\vspace{1mm}

In a recent comment \cite{kim2020}, it was claimed that the Lyapunov exponent $\lambda_{\rm L}$ of the $N$ Majoranas Sachdev-Ye-Kitaev model of strength $J$ with a quadratic perturbation of strength $\kappa$ is 
\vspace{-2mm}
\begin{equation}
\lambda_{\mathrm{L}}/\kappa = 3T^2 J^2/\kappa^4 \, \label{eq:main}
\end{equation}

\vspace{-2mm}
\noindent for a fixed and large $N$, $T/\kappa \ll 1$, $ J/\kappa \ll 1$. Therefore, according to Ref.~\cite{kim2020}, the dynamics is quantum chaotic for a sufficiently large $\kappa$ assuming the rest of parameters ($T,N,J$) are fixed.
 This is in tension with the results of Ref.~\cite{garcia2018} where we show that, for $T,N,J$ fixed, this model is only quantum chaotic for sufficiently small $\kappa$.
Here we show why the semiclassical formalism employed to obtain Eq.~(\ref{eq:main}) is not applicable in this range of parameters and show explicitly that the system is no longer quantum chaotic in the large $\kappa$ limit.
We note that the perturbative $1/N$ expansion leading to Eq.~(\ref{eq:main}) in \cite{kim2020} is valid for $t_{\rm{E}} \leq t \ll t_{\rm{H}}$ where 
$t_{\rm{H}} \sim \frac{1}{\Delta} 
$
is the Heisenberg time, $\Delta$ is the mean level spacing, $t_{\rm{E}}$ is the Ehrenfest or scrambling time and $\hbar = 1$. 
Physically, this is the time scale for which non-perturbative in $N$ quantum effects, related to the discreteness of the spectrum, become dominant. The analytical treatment of Ref.~\cite{kim2020} is semiclassical, namely, it captures only the leading $1/N$ corrections. Therefore, it is not in principle applicable when $t_{\rm{E}}$ becomes of the order or larger than $t_{\rm{H}}$. In our model, we can estimate
$
t_{\rm{H}}\sim t_0 2^{N/2},	
$
where $t_0$ is $N$ independent, 
so the formalism of Ref.~\cite{kim2020} is not applicable for

\vspace{-5mm}
\begin{align}
t_{\rm{E}} \geq t_{\rm{H}} \qquad\Leftrightarrow\qquad \lambda_{\rm{L}}(\kappa,T,J) \leq 2^{1-N/2}\log{N}.
\end{align}

\vspace{-2mm}
It is straightforward to estimate the maximum $\kappa =\kappa_{\star}$ for which Eq.~(\ref{eq:main}) applies:

\vspace{-7mm}
\begin{align}
\kappa_{\star}\left({T},N,J\right) \sim \left(\frac{TJ}{\epsilon_0^{1/2}}\right)^{2/3} \frac{2^{N/6}}{(\log{N})^{1/3}},
\end{align}

\vspace{-2mm}
\noindent where $\epsilon_0 =1/t_0$ and we omit numerical factors ${\cal O}(1)$. 

\vspace{-2mm}
\begin{figure}[htb]
	\includegraphics[scale=.54]{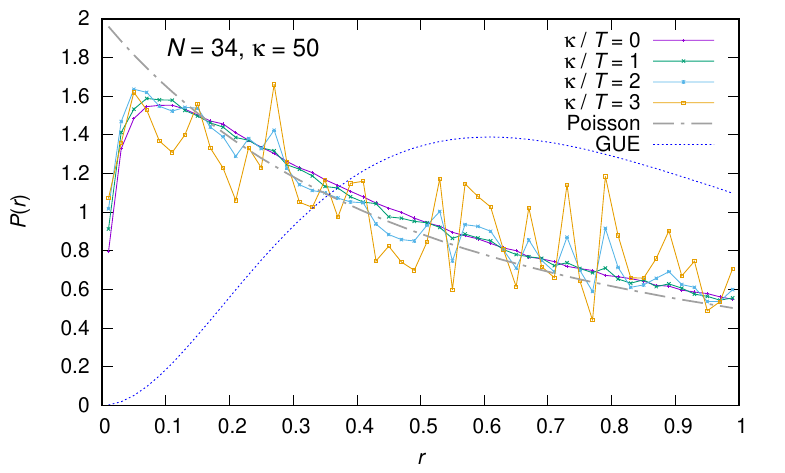}
	\includegraphics[scale=.27]{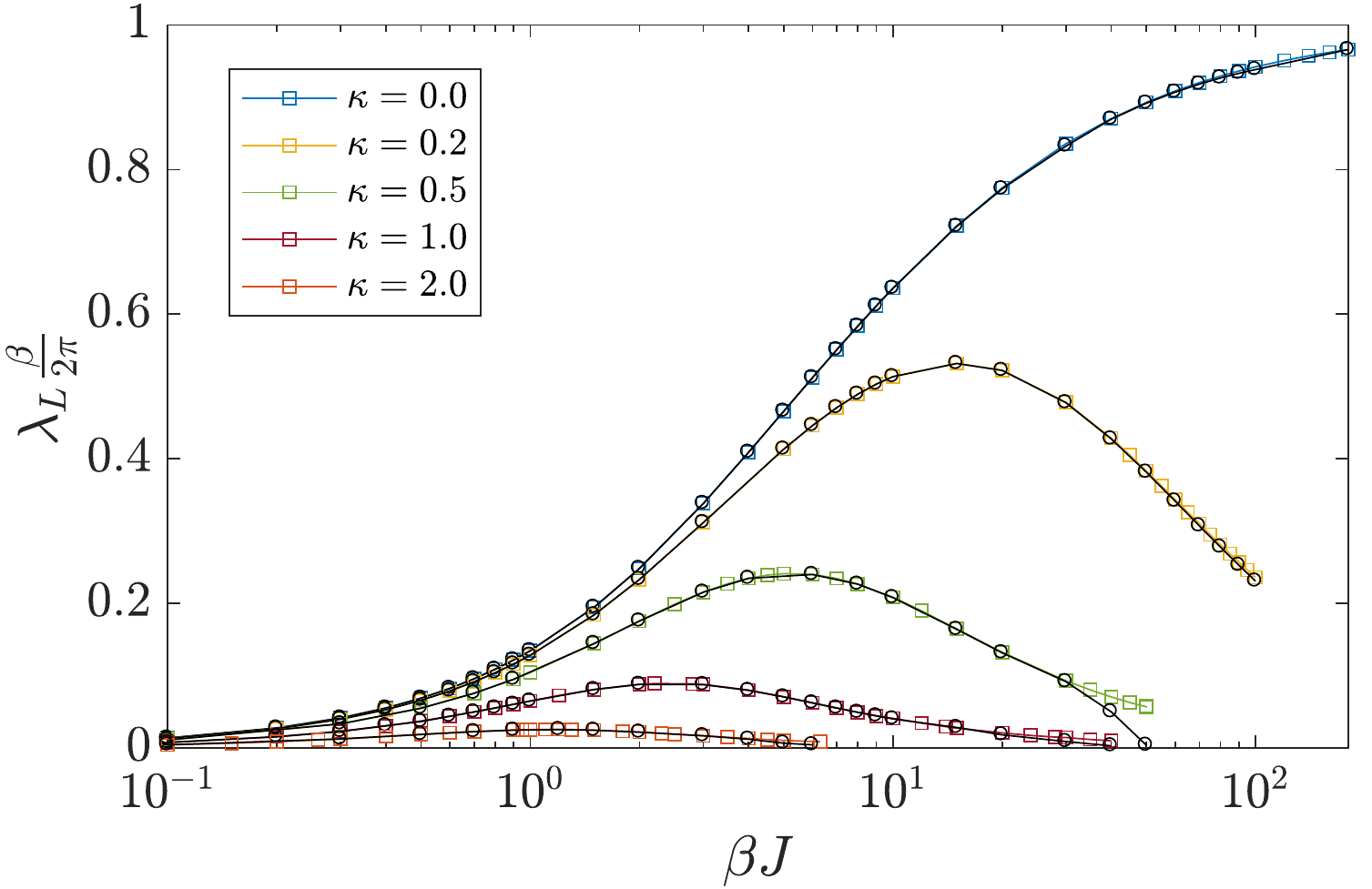}
	\vspace{-3mm}
	\caption{
Left:
Distribution function $P(r)$ of the adjacent gap ratio for $\kappa = 50$ (see definition of $r$ around Eq. 35 of Ref.~\cite{monteiro2020}), for different temperatures ($T$ is defined in Ref.~\cite{garcia2018}). Agreement with Poisson statistics contradicts the claim of Ref.~\cite{kim2020} that the system is quantum chaotic in this region.
Right:
Lyapunov exponent as a function of $\beta =1/T$ for various $\kappa$'s, circles are the results of Ref.~\cite{garcia2018} and the squares are the recalculated results.}\label{fig} 
\end{figure}

\vspace{-2mm}
This is in contradiction with the claim of Ref.\cite{kim2020} that, for any fixed value of $N \gg 1$, $T$, $J$, Eq.~(\ref{eq:main}) applies for any sufficiently large $\kappa$. The fact that this is only true for $\kappa < \kappa_{\star}$ cast doubts about the quantum chaotic nature of the motion for sufficiently large $\kappa$ which is the main claim of Ref.~\cite{kim2020}. We show next that this claim is indeed incorrect.
The fact that the system is not quantum chaotic for a fixed $J, N$ and sufficiently large $\kappa$ is rather evident from the level statistics analysis of Ref.~\cite{garcia2018}, see the adjacent gap ratio in Fig. 11 of the supplemental information of Ref.~\cite{garcia2018} 
or the lower panels of
Fig. 1 and Fig. 2 in the main paper.
As a further confirmation of this point, in the left panel of Fig.~\ref{fig}, we depict results of the distribution function of the adjacent spacing ratio $P(r)$ for $\kappa = 50$ and different temperatures. In contradiction with the claims of Ref.~\cite{kim2020} about the persistence of quantum chaos for any $\kappa/T \gg 1$, we find that, in this region, $P(r)$ is already very close to Poisson statistics that characterizes an integrable or many-body localized phase. 

Moreover, it has recently been shown analytically \cite{monteiro2020} that for $\kappa > \kappa_c \sim N^2 \log N$, the system is no longer quantum chaotic, {\it all} states are many-body localized, and spectral correlations are well described by Poisson statistics. In the language used in Ref.~\cite{kim2020}, this means that the Lyapunov exponent is zero for any temperature. The results of Ref.~\cite{garcia2018} points to a much weaker $N$ dependence, if any, of $\kappa_c$ in the tail of the spectrum corresponding to the low temperature limit. However, the analytical result of Ref.~\cite{monteiro2020} refutes already the main claim of Ref.~\cite{kim2020} that  $\lambda_\mathrm{L}$ is always finite.

Finally, 
 in the right panel of Fig.~\ref{fig}, we compare Fig. 3 of \cite{garcia2018} with the recalculated $\lambda_\mathrm{L}$ after we noticed that for a few $\kappa$'s in the low temperature limit, our code did not pick up the largest $\lambda_\mathrm{L}$. 
The corrections are only appreciable for a few points. 
 A fitting of $\lambda_\mathrm{L}(T)$ still suggests the possible existence of a critical temperature $T_c(\kappa)$ below which $\lambda_\mathrm{L} = 0$. However, the values of $T_c$ resulting from the fitting 
 are not conclusive about the vanishing of $T_c$ at finite temperature. 
 In any case, the conclusions of this reply, or of Ref.~\cite{garcia2018}, are not modified because, as argued earlier,  the theoretical formalism employed in Ref.~\cite{kim2020} breaks down for a sufficiently large $\kappa /T \gg 1$ and, more importantly, in stark contrast with the claim of Ref.~\cite{kim2020}, level statistics results together with the analytical findings of Ref.~\cite{monteiro2020}, confirm that, for a fixed $N$, the model is not quantum chaotic but many-body localized/integrable for sufficiently large $\kappa$.


\begin{minipage}{\columnwidth}%
\maketitle
\end{minipage}

\end{document}